%% file: PUB_2lepton2010_inc_noComments.tex
\RequirePackage{lineno}
\documentclass[twocolumn,epj,nopacs]{svjour}
\usepackage{atlasphysics}
\usepackage{subfigure}
\usepackage{mathrsfs}
\usepackage{graphicx}
\usepackage{multirow}
\usepackage{rotating}
\usepackage{url}
\usepackage{color}
\usepackage{setspace}
\usepackage{amsmath}
\usepackage{textpos}
\usepackage{hyperref}

\input{susydefs}

\titlerunning{Search for SUSY in events with lepton pairs and $\met$ with ATLAS}
\newcommand{\dileptitle}{Search for supersymmetric particles in events with lepton pairs and large missing transverse momentum in $\sqrt{s}=7\TeV$ proton-proton collisions with the ATLAS experiment}

\newcommand{\hideit}[1]{{ }}

\newcommand{\alpgen}{{\tt ALPGEN}}

\newcommand{\prospino}{{\tt PROSPINO}}
\newcommand{\herwig}{{\tt HERWIG}}
\newcommand{\pythia}{{\tt PYTHIA}}
\newcommand{\mcatnlo}{{\tt MC@NLO}}

\newcommand{\jimmy}{{\tt JIMMY}}

\newcommand{\ourintlumi}{{$35$~pb$^{-1}$}}
\newcommand{\btheta}{\boldsymbol \theta}

\begin{document}
\title{\dileptitle \begin{textblock}{6}[0,1](0,-3){\small CERN-PH-EP-2011-040, Submitted to EPJC Letters}
\end{textblock}
}
\author{The ATLAS Collaboration}
\institute{}
\date{}

\abstract{Results are presented of searches for the production of supersymmetric particles decaying into final states with missing transverse momentum and exactly two isolated leptons in $\sqrt{s}=7\TeV$ proton-proton collisions at the Large Hadron Collider. Search strategies requiring lepton pairs with identical sign or opposite sign electric charges are described. In a data sample corresponding to an integrated luminosity of \ourintlumi~collected with the ATLAS detector, no significant excesses are observed. 
Based on specific benchmark models, limits are placed on the squark mass between 450 and 690~GeV for squarks approximately degenerate in mass with gluinos,
depending on the supersymmetric mass hierarchy considered. }

\maketitle

Many extensions of the Standard Model (SM) predict the existence of new states decaying 
to invisible particles, often motivated by dark matter arguments. 
If such states are produced in collisions at the Large Hadron Collider, 
then they can potentially be identified by the presence of missing transverse momentum 
generated by the invisible decay products. The most important SM backgrounds, in particular jets 
from QCD production processes (referred to as ``QCD jets'' hereafter), can be suppressed by 
requiring in addition the presence of leptons in the final state. Particles predicted by 
supersymmetric (SUSY) theories \cite{susy} can be sought with such a signature, with 
the missing transverse momentum generated by the production of weakly interacting lightest supersymmetric 
particles (LSP), and the leptons produced in the cascade decay of supersymmetric particles. 

In this letter the first results of searches for the production of SUSY particles 
at ATLAS using final states with two leptons and missing transverse momentum are presented.
Leptons are produced through the decays of
charginos and neutralinos into $W$ and $Z$ bosons, and into real or virtual sleptons, the SUSY
partners of leptons, if their masses are light enough.
The main sources of leptons in SM events include $W$ and $Z$ decays,  fake leptons from misidentification of jets and non-isolated leptons from heavy flavour decays. Two search strategies are described which require, respectively, isolated leptons of 
same sign (SS) or opposite sign (OS) electrical charge. SS lepton production in SM events is rare. On the other hand, the production of 
gluinos, which decay with the same probability 
to squark$+$anti-quark and anti-squark$+$quark pairs, and of squark-squark pairs,
provides an abundant source of SS lepton pairs in SUSY events ~\cite{Barnett:likesign}.
When imposing the OS lepton pair requirement the SM background is larger. However, the signal 
cross section is also increased by the additional production of squark$+$anti-squark pairs. 
The results reported here are complementary to those from SUSY searches
requiring lepton pairs of identical flavor 
\cite{atlas-2lepfs-pub}, and also those from inclusive searches 
requiring jets, missing transverse momentum and zero leptons \cite{atlas-0lep-pub} or one 
lepton \cite{atlas-1lep-pub}. A search by CMS for SUSY in events with OS lepton pairs is reported
in Ref.~\cite{cms-2lepos-pub}.

The ATLAS detector~\cite{Aad:2008zzm} is a multipurpose particle
physics apparatus with a forward-backward symmetric cy\-lindrical
geometry and near 4$\pi$ coverage in solid angle\footnote{
ATLAS uses a right-handed coordinate system with its origin at the 
nominal interaction point (IP) in the centre of the detector 
and the $z$-axis coinciding with the axis of the beam pipe. 
The $x$-axis points from the IP to the centre of the LHC ring, 
and the $y$ axis points upward. 
Cylindrical coordinates 
$(r,\phi)$ are used in the transverse plane, $\phi$ being 
the azimuthal angle around the beam pipe. The pseudorapidity is 
defined in terms of the polar 
angle $\theta$ as $\eta=-\ln\tan(\theta/2)$.}.
The inner tracking detector (ID) consists of a
silicon pixel detector, a silicon microstrip detector (SCT), and
a transition radiation tracker (TRT). 
The ID is surrounded by a thin superconducting
solenoid providing a 2 T magnetic field, and by
high-granularity liquid-argon (LAr) sampling electromagnetic calorimeters.
A hadron calorimeter of iron-scintillator tiles provides coverage 
in the central rapidity range.
The end-cap and forward regions 
are instrumented with LAr calorimetry
for both electromagnetic and hadronic measurements.
The muon spectrometer (MS) surrounds the calorimeters  and consists 
of three large superconducting toroids, a system 
of precision tracking chambers, 
and detectors for triggering.

The full 2010 ATLAS $pp$ dataset is used in this analysis, collected at the LHC
at a centre-of-mass energy of 7 \TeV. 
Application of basic beam, detector and data-quality requirements results in
a dataset corresponding to a total integrated luminosity of \ourintlumi. The uncertainty on the
integrated luminosity is estimated to be 11\%~\cite{atlas-lumi}.
The data have been collected with a single lepton ($e$ or $\mu$) trigger. The detailed
trigger requirements vary throughout the data-taking period owing to the rapidly
increasing LHC luminosity and the commissioning of the trigger system. The requirements are such that the trigger efficiency is constant and stable for leptons with transverse momentum \pt$>$20~GeV. 
The efficiency of the triggers has been studied using data, 
and agrees well with expectations.

Monte Carlo (MC) event samples are used to develop and validate the 
analysis procedure, determine detector acceptance and reconstruction efficiency, 
and transfer background expectations from control regions to signal regions. These 
samples are also used to model the sub-dominant SM backgrounds. Samples of 
QCD jet events are produced with the \pythia~ generator~\cite{pythia}.
Production of top quark pairs and single top is simulated with the \mcatnlo~ generator~\cite{mcatnlo},
with an assumed top-quark mass of $172.5$ \GeV.
Samples of $W$ and $Z/\gamma^*$ production with accompanying jets are produced
with the \alpgen~ generator~\cite{alpgen} for $m_{\ell\ell}>40$~GeV. Low mass dileptons from $Z/\gamma^*$ production
are generated with \pythia. Di-boson ($WW$, $WZ$, $ZZ$) production is simulated
with the \herwig~ generator~\cite{herwig,herwig65}. 
All the MC samples are normalised to the available 
next-to-next-to-leading order (NNLO) or next-to-leading order (NLO) 
QCD calculations, except the QCD jet sample, which is normalised to the leading order \pythia\ cross section.
Fragmentation and had\-ronization for the \alpgen~ \newline and \mcatnlo~ 
samples is performed with \herwig, using
\jimmy~\cite{Butterworth:1996zw} for the underlying event model. The MC samples are produced using the ATLAS detector simulation software
~\cite{mc09tunes} based on {\tt GEANT4}~\cite{geant4}. The MC samples are tuned to
reproduce the same average number of primary vertices as in the data in order
to take into account multiple inelastic interactions in the same beam crossing.

Criteria for electron and muon identification closely follow those
described in Ref.~\cite{atlas-winclusive}.
Electrons in the signal region are required to pass the ``tight'' selection
criteria, have $\pt > 20~\GeV \,$ and $|\eta| <~2.47$.
Events are removed if an electron satisfying the ``medium'' selection is found in the transition region between the barrel and end-cap electromagnetic calorimeter, $1.37 < |\eta| <~1.52$. The medium criteria are mainly based on 
lateral shower shape requirements in the calorimeter, 
$E/p$ (where $E$ is the shower energy in the calorimeter and $p$ the track momentum
in the ID) and TRT cuts are also applied for the tight electron selection, which 
provides additional rejection against conversions and fakes from hadrons.
Muons are required to be identified either in both the ID
and MS systems (combined muons)  or as a match
between an extrapolated ID track and one or more segments in the MS.
For combined muons, a good match between ID and MS tracks is required, and
the $\pt$ values measured by these two systems
must be compatible within the resolution. The summed $\pt$ of other ID tracks with \pt$>$ 500 \MeV~
within a distance $\Delta R = \sqrt{(\Delta \eta)^2 + (\Delta \phi)^2} < 0.2$ 
around the muon track is required to be less than $1.8$ \GeV.
Only muons with $\pt > 20 \GeV$ and $|\eta| < 2.4$ are considered.
For the final selection, the distance between the $z$ coordinate of the 
primary vertex and that of the extrapolated muon track at the point
of closest approach to the primary vertex
must be less than $10$ mm.

Jets are reconstructed using the anti-$k_T$ jet clustering 
algorithm~\cite{anti-kt} with a distance parameter $R = 0.4$. 
They are corrected for calorimeter non-compensation, upstream material
and other effects using $\pt$ and $\eta$ dependent calibration
factors obtained from Monte Carlo and validated with extensive
test-beam and collision-data studies~\cite{atlascalibration}. 
Only jets with $\pt > 20~\GeV \,$
and $|\eta| < 2.5$ are considered.
If a jet and a selected electron overlap within a distance
$\Delta R < 0.2$, the jet is discarded.
Furthermore, identified medium electrons or muons are considered only if they 
satisfy $\Delta R > 0.4$ with respect to the closest remaining jet.
Events are discarded if they contain any jet failing
basic quality selection criteria that reject detector noise and 
non-collision backgrounds~\cite{atlas-jet-clean}. 
The calculation of missing transverse momentum ($\met$) is based on the
modulus of the vector sum of the transverse momenta of the reconstructed objects (jets
with $\pt > 20 \GeV$ over the full calorimeter coverage $|\eta| < 4.9$ and selected leptons), together with any additional non-isolated muons and
calorimeter clusters not belonging to reconstructed objects.

Events failing the requirement of at least one reconstructed
primary vertex with at least five associated tracks are rejected.
Selected events must contain exactly two leptons ($e$ or $\mu$) after the object
selection described above. For electrons an isolation criterion is required: the
summed calorimeter transverse energy within a distance $\Delta R < 0.2$ around the electron
divided by the \pt\ of the electron must be smaller than 0.15.
The invariant mass ($m_{\ell\ell}$) of the lepton pair must be greater than 5~GeV. 
The signal region for OS (SS) events is defined by the requirement $\MET$ $>$ 150~GeV (100~GeV), which was chosen through 
optimisation of the expected signal significance
for a selection of models drawn from the Minimal Supersymmetric Model (MSSM) framework in a mass range just above the existing limits from direct searches.

The main background for the SS analysis arises from SM 
processes generating events containing at least one fake or non-isolated 
lepton. These processes are collectively referred to as ``fake lepton'' background, and mainly consist of $t\bar{t}$, single-top, $W+$jets and QCD light and heavy flavour jet production.
The other significant backgrounds arise from di-boson production and from charge mis-measu\-rements of electrons in $t\bar{t}$ events 
that have undergone hard bremsstrahlung with subsequent photon conversions. The other SM backgrounds, such as $Z$ boson production, are small, since their contribution is largely suppressed by the $\met$ cut.
For the OS analysis the dominant background arises from $t\bar{t}$ production.
In addition, there are contributions from fake or non-isolated leptons, $Z$+jet, di-boson and single-top production.

For the ``fake lepton'' background, the origin of
the detected leptons are either jets faking leptons or 
heavy flavoured meson decays into non-isolated leptons. 
The contribution from this background is estimated from data using a method that
is similar to that described in Ref.~\cite{atlas-topxsec}. This method defines a 
looser lepton selection, referred to as ``loose'' hereafter, and counts the numbers of observed events containing loose-loose, loose-tight, tight-loose and tight-tight lepton pairs. The probability of loose real leptons to pass the tight selection criteria is obtained using a $Z\to \ell^+\ell^-$ control sample while the probability of loose fake leptons to pass the tight selection criteria is obtained using several control samples dominated by QCD jet events.
Using these probabilities,
linear equations can be obtained for the observed event counts as functions of the numbers of events containing fake-fake, fake-real, real-fake and real-real lepton pairs. These four equations can be solved simultaneously to yield the fake lepton background for the SS and OS analyses.

The contribution from the incorrect electron charge assignment background to the SS analysis 
is studied using $Z \to e^+e^-$ MC events by comparing the charges of
generator level electrons to those of reconstructed electron candidates following the application of
the SS analysis cuts. The background contribution is calculated as a function of the electron rapidity and applied to
$t\bar{t} \to e^{\pm}\ell^{\mp}$ ($\ell=e,\mu$) MC events to obtain the $t\bar{t}$ contribution in the SS analysis.
The method is validated with data by looking at the number of SS $Z \to e^{\pm}e^{\pm}$ events in a 
sample selected by requiring a lepton pair with invariant mass between 60 GeV and 120 GeV. The method predicts 61.3 $\pm$ 0.4 events 
compared with 62 observed events in data.

The number of $t\bar{t}$ events in the OS signal region (SR) is obtained by 
multiplying the observed number of $t\bar{t}$  events in an appropriately defined control region (CR)
by a factor $F(CR\rightarrow SR)$, defined as the ratio between the number
of $t\bar{t}$ MC events in the SR and the number of MC events in the CR. 
A $t\bar{t}$ dominated control region is defined by selecting ``top-tagged'' lepton pair events
which satisfy the same selection criteria as signal candidates except for a
$60<\MET<80$~GeV requirement, defining a region in which both the $Z$ contribution and the SUSY signal contamination are small.
Events in this region are top-tagged using the variable $\mct$, introduced in Ref.~\cite{Tovey:2008ui}.
For two identical decays of heavy particles into two visible
particles (or particle aggregates) $v_{1}$ and $v_{2}$, and
into invisible particles, $\mct$ is defined as:
\begin{equation}
  \mct^{2}(v_{1},v_{2}) = \left[ E_{\rm T}(v_{1}) + E_{\rm T}(v_{2}) \right]^2 -  \left[  {\bf p_T}(v_{1}) - {\bf p_T}(v_{2}) \right]^2,
  \label{eq:mct_endpoints}
\end{equation}
\begin{figure}
  \begin{center}
    \includegraphics[width=0.45\textwidth]{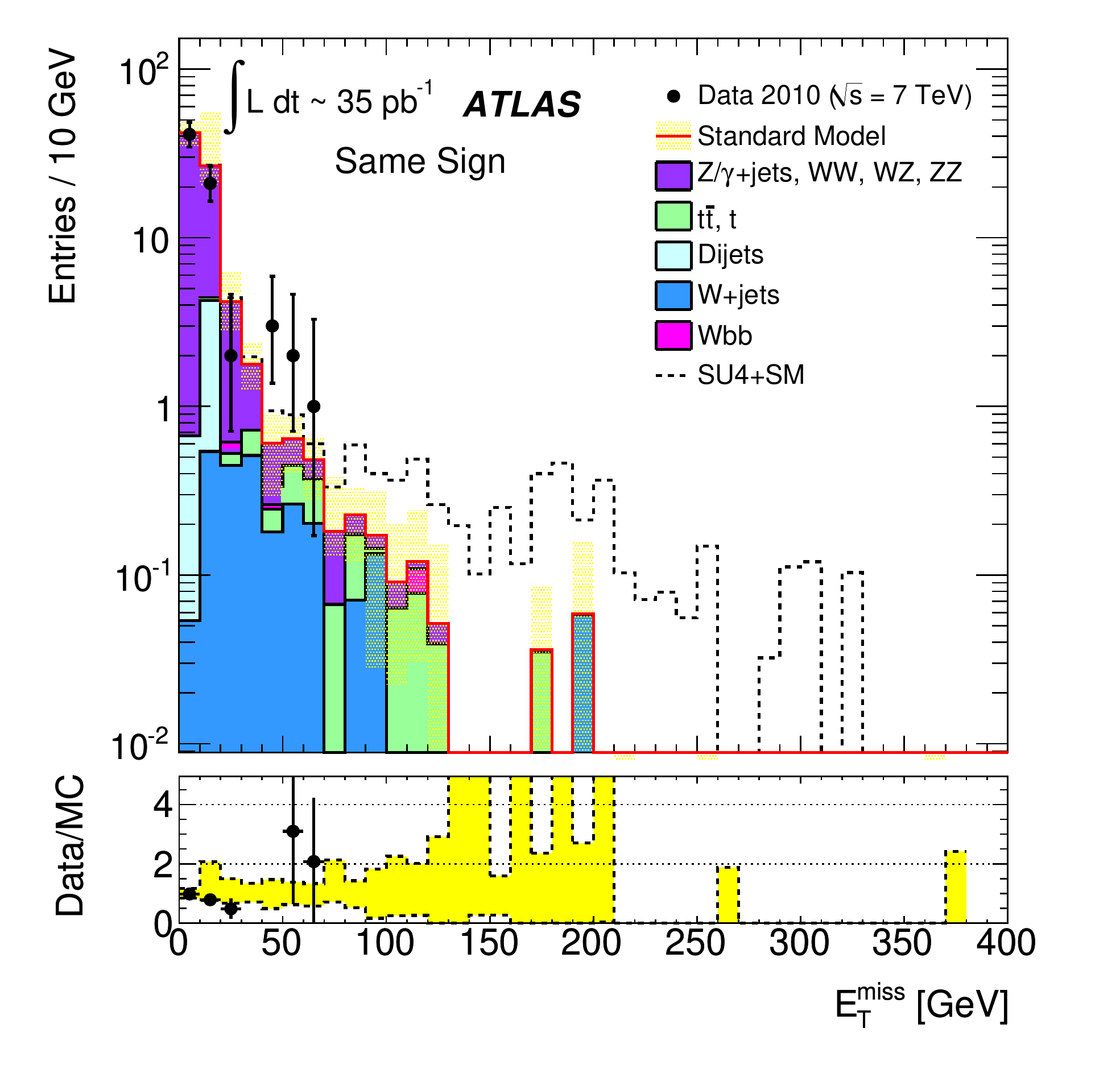}
    \includegraphics[width=0.45\textwidth]{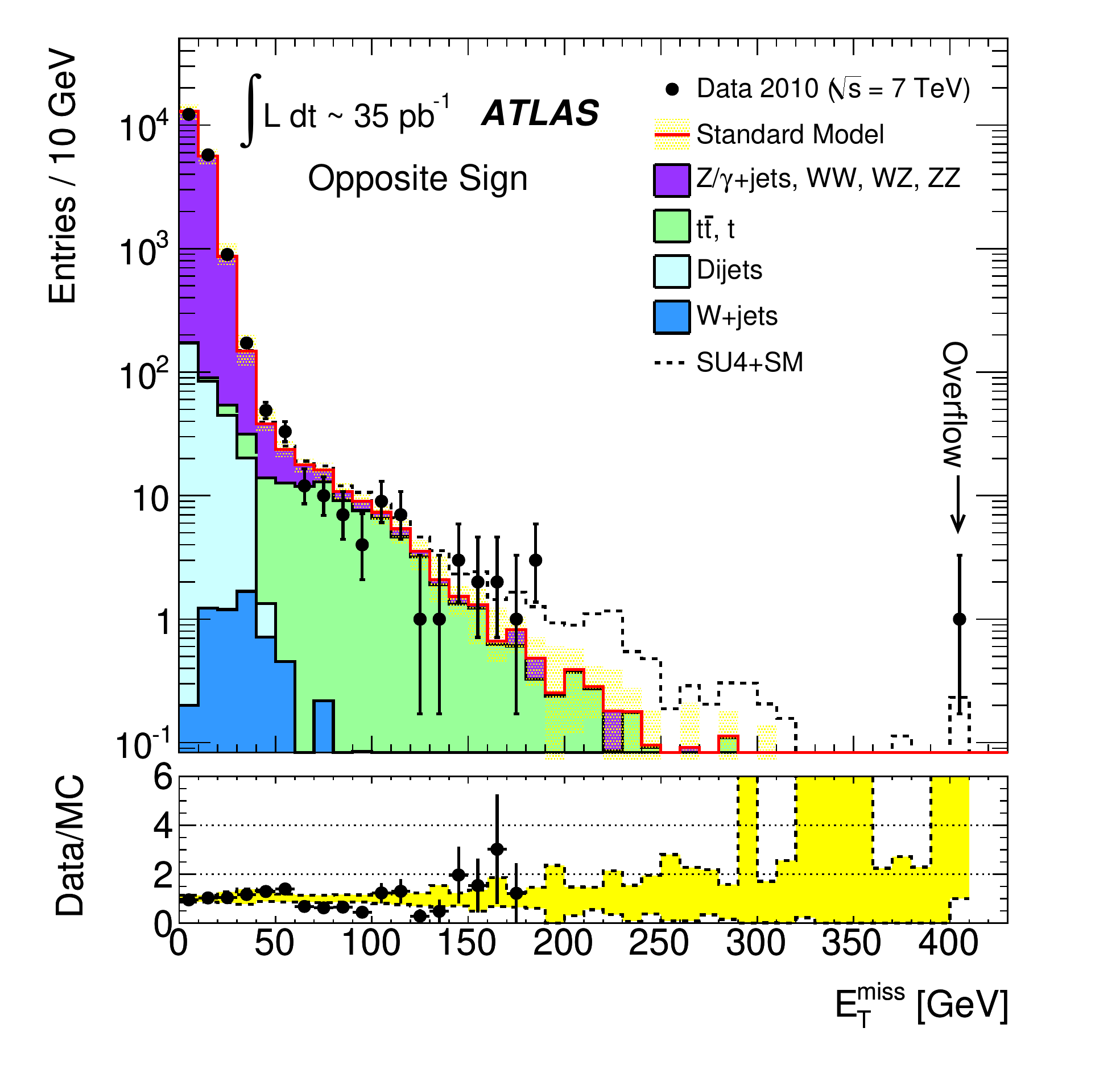}
    \caption{\label{Etmissfig} Distributions of $\MET$ for SS (upper) and OS (lower) lepton pair events passing the analysis selections. 
      The data are shown as points with error bars superimposed on the expected SM background distributions 
      determined (mostly data-driven) with MC simulation. The overflow point in the OS $\MET$ histogram is a 
      likely candidate for cosmic ray interaction. 
      In the bottom panel the ratio between data and the total SM background is shown.
      The histogram labelled ``Standard Model'' represents the sum of all backgrounds and the 
      light (yellow) bands indicate the uncertainty on the MC predictions from 
      finite MC statistics and uncertainties in cross section, luminosity and jet and lepton energy scales and 
      resolutions. ``SU4'' represents a point in the mSUGRA/CMSSM parameter space
      with $m_0=200$ GeV, $m_{1/2}=160$ GeV, $A_0=-400$ GeV, $\tan\beta=10$ and $\mu >$0. }
  \end{center}
  \vspace*{-1.1cm}
\end{figure}
where transverse momentum vectors are denoted by ${\bf p_{T}}$ and transverse energies $E_{\rm T}$ are defined as $E_{\rm T} = \sqrt{p_{\rm T}^2 +m^2}$. In Equation~\ref{eq:mct_endpoints} $v_i$ can be a lepton, a jet, or a lepton-jet combination. The distributions of $\mct$  for each of these combinations,
as well as the distributions of invariant mass for jet$+$lepton pairs generated in the
same top quark decay, possess kinematic end-points which are functions of the masses of the top quark and $W$ boson
as detailed in Ref.~\cite{Polesello:2009rn}.
An event is considered to be top-tagged
if it includes two jets with \pt$>20$~GeV and the three $\mct$
variables and the lepton-jet invariant masses are compatible with the kinematics of fully leptonic $t\bar{t}$
($t\bar{t} \to \ell^+ \nu \ell^- \bar{\nu} \bbbar$) events.
A total of  15 top-tagged data events are observed in the CR compared with a MC expectation of 
$21.3\pm3.8$ events, of which 18.8 arise from $t\bar{t}$ production and 2.5 from 
other SM sources. The quoted uncertainty 
is the statistical error on the MC samples.  
The estimated total number of $t\bar{t}$ events in the SR 
for the OS analysis is $2.9^{+1.4}_{-1.3}$. The quoted uncertainties 
include the statistical
uncertainty on the number of events observed in the CR and the 
systematic uncertainty on the MC extrapolation to high $\met$. 
The latter arise from MC modeling of 
top quark production and decay (23\%),  and uncertainties 
in jet energy scale~\cite{atlas-inclusivejet} and 
resolution~\cite{atlas-jer} (23\%). 
The sources of uncertainty in the Monte Carlo modeling considered were 
the choice of the NLO generator, the choice of the parton shower, 
and the modeling of initial and final state QCD radiation.  
The resulting  total systematic uncertainty
on the estimated number of $t\bar{t}$ events in the OS analysis is  44\%.
The predicted $t\bar{t}$ background contribution
in the SR for the OS analysis, broken down into the three possible 
lepton flavour combinations, is given in Table~\ref{tab:expectedEvents}.

A partially data-driven approach is adopted to estimate the contribution 
from $Z$ production in the $e^+e^-$ and $\mu^+\mu^-$ channels of the OS analysis. 
A control region is defined requiring $\met<20$~GeV and 
$81<m_{\ell\ell}<101$~GeV, where non-$Z$ contributions
are found to be negligible. A normalisation factor between the CR and the SR 
is obtained from this region using the MC. This factor is applied to the data in the CR in order to estimate the
contributions to the signal regions.
The $e\mu$ contribution is estimated solely using MC due to
lack of events in the control region. The resulting numbers
are presented in Table~\ref{tab:expectedEvents}. The contributions from other SM processes such as single-top and 
di-boson production are estimated using MC samples and found to be small. 
The contribution of events from cosmic ray interactions is
considered only for the $e\mu$ channels (arising from a single cosmic ray muon in coincidence with a
collision electron) and the opposite sign $\mu\mu$ channels (arising from a single
cosmic ray muon traversing the detector which is reconstructed as two
opposite sign muons). This contribution is extracted from the number of selected muons which
fail a tight cut on the minimum distance in the transverse plane of the associated inner 
detector track from the reconstructed primary vertex. The method requires knowledge of the
probabilities for cosmic and collision muons to fail this cut. The former
is measured from a dedicated data sample selected with a cosmic ray
trigger, while the latter is extracted from simulation.

The observed numbers of events and the expected numbers of SM background events 
in the signal regions for the SS  and OS analyses
are shown in Table~\ref{tab:expectedEvents}. The expected total number of SM events 
is 0.28 $\pm$ 0.14 for the SS analysis compared with zero events observed in the data,
and $3.7 \pm 1.6$ for the OS analysis compared with 9 events observed in the data. 
The $\MET$ distributions measured with data for both analyses and the expectations from Monte Carlo simulation of Standard Model processes are shown in Figure~\ref{Etmissfig}. For the SS channel the background expectations are found to be in agreement with the observation.
All of the 9 selected OS events were visually scanned, and the highest $\MET$ $\mu\mu$ event ($\sim 600 \GeV$, in overflow in Figure~\ref{Etmissfig}) was found to be a likely candidate for a cosmic ray interaction in the detector.
The number of observed events in the OS analysis is larger but in reasonable agreement 
with the background expectation. The channels with the most significant deviation are 
$e\mu$ and $\mu\mu$, for which the probabilities of the background to exceed the 
number of observed events are 12\% and 13\% respectively. The combined probability of the
$ee$, $e\mu$ and $\mu\mu$ channels is 12.8\%.

\begin{table}[!htbp] \small
  \begin{center}
   \caption{Total number of observed events in the SS and OS signal regions together with
    background expectations for an integrated luminosity
   of \ourintlumi. 
   The negative numbers for the fakes and the cosmics are an artifact of the matrix method and are taken as zero
   when calculating the total number of background events.}
  \begin{tabular}{lccc} \hline \hline
    \multicolumn{4}{c}{Same Sign, $\met>100$~GeV}\\ \hline
    & $e^{\pm}e^{\pm}$ &$e^{\pm}\mu^{\pm}$ & $\mu^{\pm}\mu^{\pm}$
    \\\hline 
  Data               & 0 &  0 &  0 \\\hline
  Fakes              & 0.12 $\pm$ 0.13  &  0.030 $\pm$ 0.026 & 0.014 $\pm$ 0.010 \\
  Di-bosons          & 0.015 $\pm$ 0.005 & 0.035 $\pm$ 0.012  & 0.021 $\pm$ 0.009 \\
  Charge-flip        & 0.019 $\pm$ 0.008 & 0.026 $\pm$ 0.011  & - \\
  Cosmics            & -     & $0^{+1.17}_{-0}$ & - \\
  Total              & 0.15 $\pm$ 0.13 & 0.09 $^{+1.17}_{-0.03}$ & 0.04 $\pm$ 0.01 \\
\hline
\hline
\multicolumn{4}{c}{Opposite Sign, $\met>150$~GeV}\\
\hline
      & $e^+e^-$ &$e^{\pm}\mu^{\mp}$ & $\mu^+\mu^-$
                 \\\hline 
  Data               & 1 &  4 &  4 \\\hline
  $t\bar t$          &$0.62^{+0.31}_{-0.28}$
                     &$1.24^{+0.62}_{-0.56}$
                     &$1.00^{+0.50}_{-0.45}$ \\
  Z+jets             &$0.19 \pm 0.15$ & $0.08 \pm 0.08$ & $0.14 \pm 0.17$ \\
  Fakes              &$-0.02\pm 0.02$ & $-0.05\pm 0.04$ & - \\
  Single top         &$0.03 \pm 0.05$ & $0.06 \pm 0.08$ & $0.10 \pm 0.07$ \\
  Di-bosons          &$0.09 \pm 0.03$ & $0.06 \pm 0.03$ & $0.15 \pm 0.03$ \\
  Cosmics            &       -        & $-0.2 \pm 1.18$ & $-0.43 \pm 1.27$ \\
  Total              & $0.92^{+0.42}_{-0.40}$
                     & $1.43^{+1.45}_{-0.59}$
                     & $1.39^{+1.41}_{-0.53}$ \\
    \hline
    \end{tabular}
    \label{tab:expectedEvents}
  \end{center}
 
\end{table}

Limits are set on the contributions to the considered final states from new physics. 
The likelihood function used to fit the event counts in the signal
regions can be written as $L(n|s,b,{\btheta})=P_S \times C_{\rm syst}$, 
where $n$ represents the number of observed data events, $s$ is the new physics signal to be
tested, $b$ is the background and ${\btheta}$ represents the systematic uncertainties, which are 
treated as nuisance parameters with Gaussian probability density functions. 
$P_S$ is the Poisson probability distribution for the event count in the signal region and $C_{\rm syst}$ represents
the constraints on systematic uncertainties taking into account correlations.
The limits are then derived from the profile likelihood ratio,
$\Lambda(s) = -2(\ln L(n|s,\hat{\hat{b}},\hat{\hat{\btheta}}) -\ln L(n|\hat{s},\hat{b},\hat{\btheta}))$, 
where $\hat{s}$, $\hat{b}$ and $\hat{\btheta}$ maximise the likelihood
function and $\hat{\hat{b}}$ and $\hat{\hat{\btheta}}$ maximise the likelihood for
a given choice of $s$. Only signal hypotheses which lead to a non-negative number of observed events are considered.
Exclusion $p$-values are obtained using pseudo-experiments with test statistic $\Lambda(s)$
and one-sided upper limits set \cite{higgscombi}.
Using the observed numbers of data events and background expectations in the signal region,
95\% confidence upper limits on the cross section times branching ratio times acceptance times efficiency
are obtained for new physics processes producing lepton pairs and $\MET$
of 0.07~pb (SS channels), 0.09~pb ($e^+e^-$ channel), 0.21~pb ($\mu^+\mu^-$ channel) and 
0.22~pb ($e^{\pm}\mu^{\mp}$ channel). 
For the SS analysis the limits are calculated using the sum of the three different channels $ee$, $\mu\mu$ and $e\mu$.
For the OS analysis limits are calculated for the three channels separately, and then combined statistically, 
as in SUSY models the signal resulting in OS same-flavour pairs  may be different
from the one generating different-flavour pairs.

\begin{figure}
\begin{center}
\includegraphics[width=0.45\textwidth]{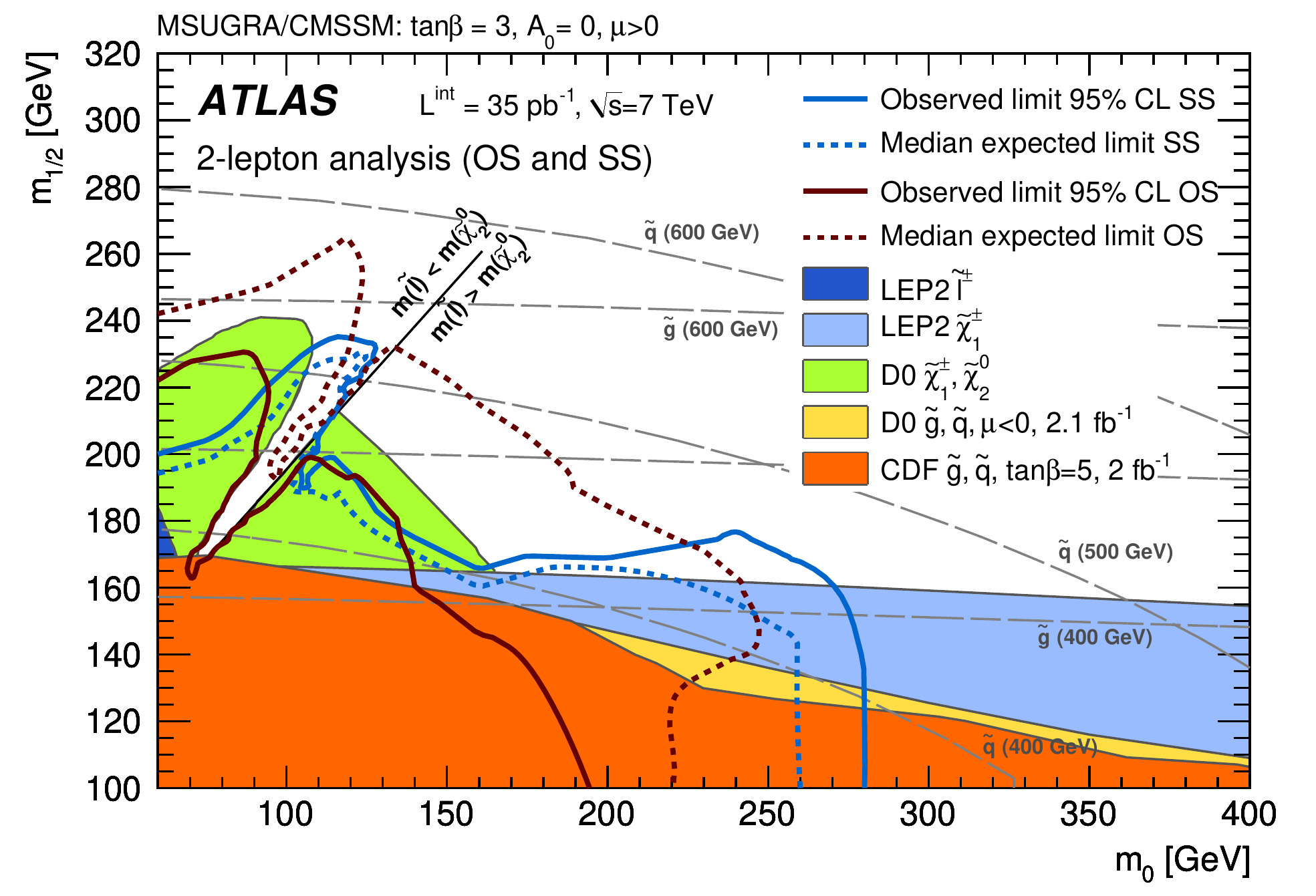}
\caption{\label{fig:sugraExclusion_m0m12}Exclusion in the mSUGRA/CMSSM \cite{msugra,cmssm} $(m_0,m_{1/2})$ plane
for $\tan\beta=3$, $A_0=0$ and $\mu>0$, together with existing limits \cite{cdfsquarksgluinos,d0squarksgluinos,d0trilepton,lepsusy}. The expected (dashed line) and observed (full line) 
95\% C.L. exclusion limits are shown for the opposite sign (black line) 
and same-sign (blue line) analyses. The illustrated D0 limit assumes $\mu<0$.}\end{center}
\vspace*{-0.5cm}
\end{figure}

Within the mSUGRA/CMSSM framework \cite{msugra,cmssm}, these results are interpreted as limits in the $(m_0,m_{1/2})$ plane, for the $\tan\beta=3$, $A_0=0$, $\mu>0$ slice of the model. Model grids in a more general MSSM 24-parameter framework 
as defined in Ref.~\cite{Paige:2003mg} are also studied. For these models (referred to as ``MSSM PhenoGrid2'' hereafter) the following parameters are fixed: $m_A=1000$ GeV, $\mu = 1.5\times\min(m_{\tilde{g}},m_{\tilde{q}})$,  $\tan{\beta}=4$,  
$A_t$=$\mu/\tan{\beta}$, $A_b=\mu\tan{\beta}$, and $A_l=\mu\tan{\beta}$.
The masses of third generation sfermions are set to 2~TeV, and 
common squark and slepton mass parameters are assumed for the first 
two generations. The remaining free parameters are 
the three gaugino masses and the squark and slepton masses. Two grids in the $(m_{\tilde{g}},m_{\tilde{q}})$ plane are generated: 
one yielding soft final state kinematics, defined by
$m_{\tilde{\chi}_2^0}=M-50$ GeV, $m_{\tilde{\chi}_1^0}=M-150$ GeV and $m_{\tilde{l}_L}=M-100$ GeV, where $M$ 
is the minimum of the gluino and squark mass (``compressed spectrum'' models); and one with a 
very light LSP, yielding a harder spectrum of leptons, jets and $\met$, with $m_{\tilde{\chi}_2^0}=M-100$ GeV,
$m_{\tilde{\chi}_1^0}=100$ GeV and $m_{\tilde{l}_L}=M/2$ GeV (``light neutralino'' models). SUSY signal events are generated with {\tt \herwig++}~\cite{herwigpp}
for the mSUGRA/CMSSM models and with \herwig ~for the MSSM models. Cross sections are calculated
at NLO with \prospino~\cite{prospino}. Theoretical and experimental uncertainties 
on the signal rate are calculated for each model. Theoretical 
uncertainties are evaluated by varying the factorisation and renormalisation
scales and by varying the {\tt CTEQ6.6} PDF sets~\cite{cteq66} used for the cross section
calculation. Experimental uncertainties include those due to the lepton and jet energy scale and resolution and 
an 11\% uncertainty on the luminosity measurement. The total uncertainty varies between 20\% and 30\% for most of the
signal models considered in this analysis. 

\begin{figure}
\begin{center}
\includegraphics[width=0.45\textwidth]{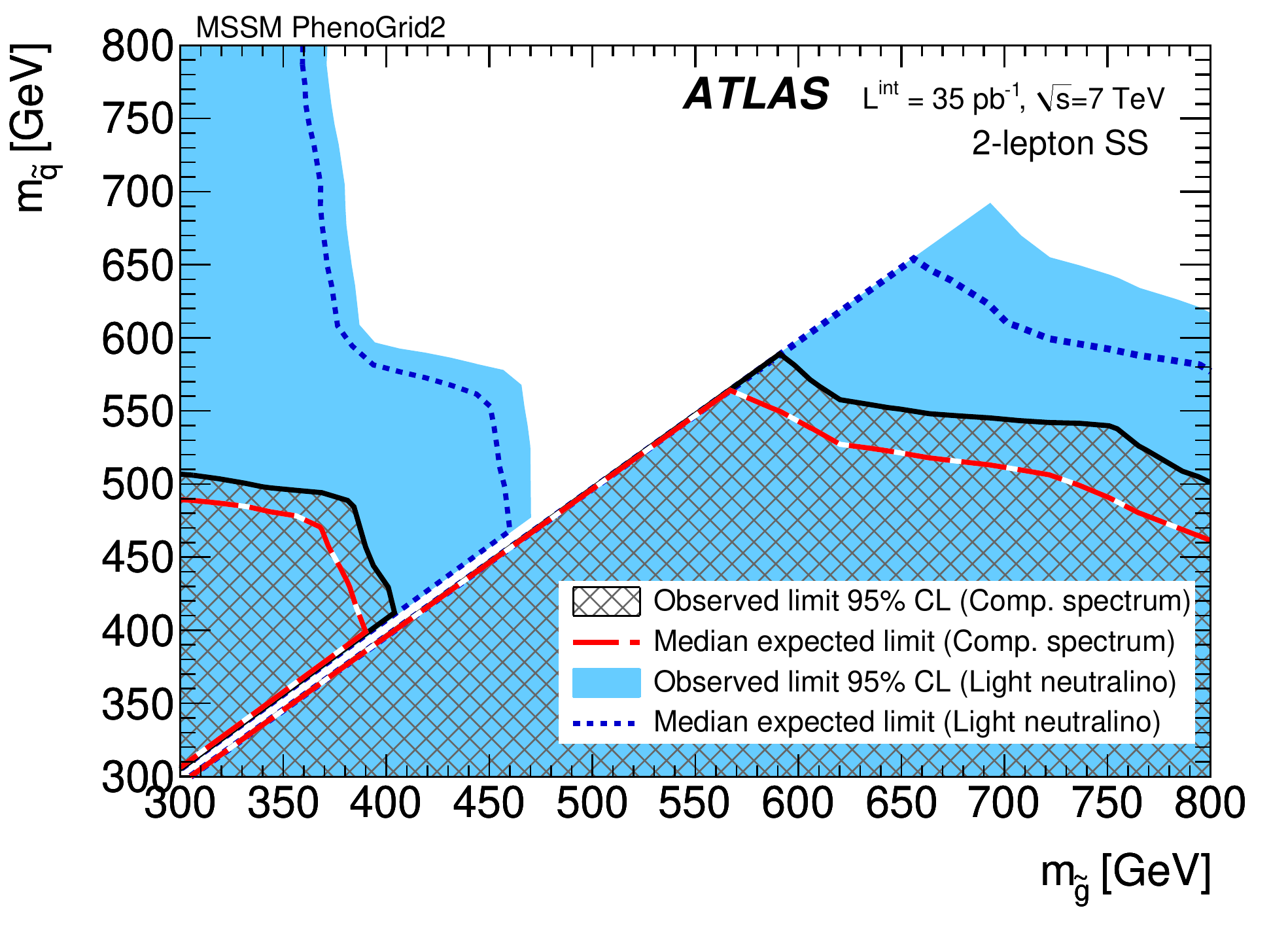}
\includegraphics[width=0.45\textwidth]{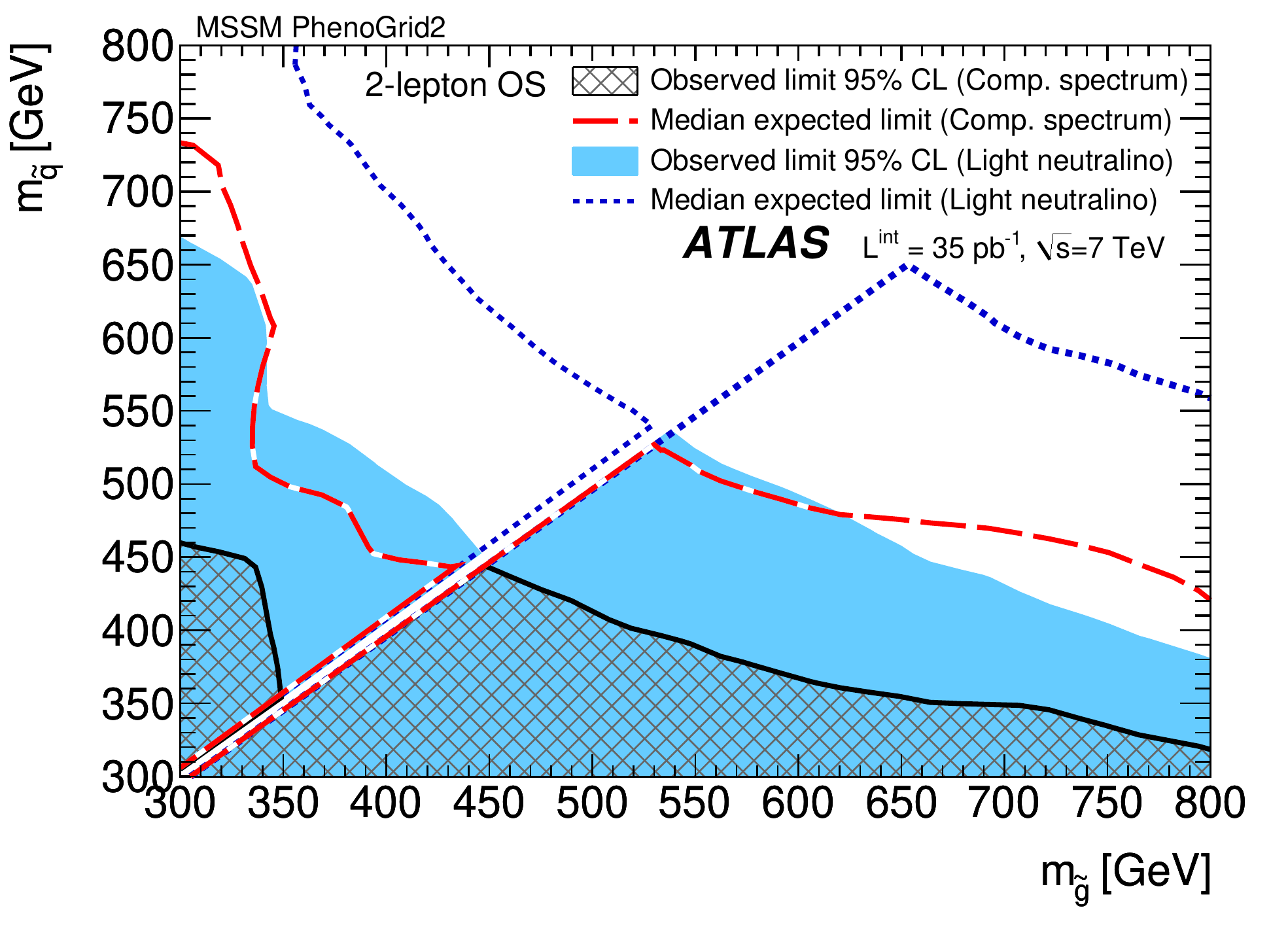}
\caption{\label{FinalFSLimits}Expected and observed 95\% C.L. exclusion limits in the $(m_{\tilde{g}},m_{\tilde{q}})$ plane for the 
specific MSSM models described in the text.
The upper panel is for the SS analysis, the lower panel for the OS analysis.
}\end{center}
\vspace*{-0.5cm}
\end{figure}

The expected and observed limits in the $(m_0,m_{1/2})$ mSUGRA/CMSSM plane are shown in 
Figure~\ref{fig:sugraExclusion_m0m12} for both the OS and SS analyses. 
The excluded region of parameter space is similar to that excluded by the Tevatron experiments based on the study of 
trilepton final states \cite{d0trilepton}, and exceeds the Tevatron squark and gluino mass limits from signatures 
including jets and $\met$ \cite{cdfsquarksgluinos,d0squarksgluinos}.
For the MSSM grids the results are shown in the $(m_{\tilde{g}},m_{\tilde{q}})$
plane in Figure~\ref{FinalFSLimits} for the SS analysis (upper panel) and
OS analysis (lower panel). 
For the considered models and $m_{\tilde{g}}=m_{\tilde{q}}+10$~GeV, 
the lower limits on the squark mass for the ``compressed spectrum'' (``light neutralino'') scenarios
are 450 (550) GeV and 590 (690) GeV for the OS and SS analysis respectively.  
The achieved limits extend the region of  squark and gluino mass explored with direct searches based on jets and $\MET$ by previous experiments.

In conclusion, 
a search for the production of SUSY particles giving rise to final state with a pair of leptons and large $\MET$ has been carried out using \ourintlumi~of data collected by the ATLAS experiment at the LHC in 2010. Two analyses have been performed, using respectively same sign and opposite sign lepton pair signatures.
The observed numbers of events in the signal regions of both analyses are compatible with SM expectations. These results have been interpreted as limits in the parameter spaces of three different SUSY models, namely the mSUGRA/CMSSM framework, and two classes of MSSM models with a compressed SUSY particle mass spectrum and with a light neutralino. 
Depending on model assumptions  squarks with masses between 450 and 690~GeV 
are excluded, for squarks approximately mass degenerate and
ligher than gluinos, extending the coverage of previous experiments.

\input{acknowledgement}

\bibliographystyle{atlasnote}
\bibliography{PUB_2lepton2010_inc_noComments}
\newpage
\input{atlas_authlist_21March}

\end{document}

%% file: susydefs.tex
\newcommand{\mct}{\ensuremath{m_\mathrm{CT}}}

\def\lsim{\mathrel{\rlap{\lower4pt\hbox{\hskip1pt$\sim$}}
    \raise1pt\hbox{$<$}}}                
\def\gsim{\mathrel{\rlap{\lower4pt\hbox{\hskip1pt$\sim$}}
    \raise1pt\hbox{$>$}}}                

\makeatletter
\renewcommand{\p@subfigure}{\arabic{figure}}
\renewcommand{\thesubfigure}{\alph{subfigure}}
\renewcommand{\@thesubfigure}{(\thesubfigure)}
\makeatother

%% file: acknowledgement.tex
We thank CERN for the very successful operation of the LHC, as well as the
support staff from our institutions without whom ATLAS could not be
operated efficiently.

We acknowledge the support of ANPCyT, Argentina; YerPhI, Armenia; ARC,
Australia; BMWF, Austria; ANAS, Azerbaijan; SSTC, Belarus; CNPq and FAPESP,
Brazil; NSERC, NRC and CFI, Canada; CERN; CONICYT, Chile; CAS, MOST and
NSFC, China; COLCIENCIAS, Colombia; MSMT CR, MPO CR and VSC CR, Czech
Republic; DNRF, DNSRC and Lundbeck Foundation, Denmark; ARTEMIS, European
Union; IN2P3-CNRS, CEA-DSM/\-IRFU, France; GNAS, Georgia; BMBF, DFG, HGF, MPG
and AvH Foundation, Germany; GSRT, Greece; ISF, MINERVA, GIF, DIP and
Benoziyo Center, Israel; INFN, Italy; MEXT and JSPS, Japan; CNRST, Morocco;
FOM and NWO, Netherlands; RCN, Norway; MNiSW, Poland; GRI\-CES and FCT,
Portugal; MERYS (MECTS), Romania; MES of Russia and ROSATOM, Russian
Federation; JINR; MSTD, Serbia; MSSR, Slovakia; ARRS and MVZT, Slovenia;
DST/\-NRF, South Africa; MICINN, Spain; SRC and Wallenberg Foundation,
Sweden; SER, SNSF and Cantons of Bern and Geneva, Switzerland; NSC, Taiwan;
TAEK, Turkey; STFC, the Royal Society and Leverhulme Trust, United Kingdom;
DOE and NSF, United States of America.

The crucial computing support from all WLCG partners is acknowledged
gratefully, in particular from CERN and the ATLAS Tier-1 facilities at
TRIUMF (Canada), NDGF (Denmark, Norway, Sweden), CC-IN2P3 (France),
KIT/GridKA (Germany), INFN-CNAF (Italy), NL-T1 \-(Netherlands), PIC (Spain),
ASGC (Taiwan), RAL (UK) and BNL (USA) and in the Tier-2 facilities
worldwide.

%% file: atlas_authlist_21March.tex
\begin{flushleft}
{\Large The ATLAS Collaboration}

\bigskip

G.~Aad$^{\rm 48}$,
B.~Abbott$^{\rm 111}$,
J.~Abdallah$^{\rm 11}$,
A.A.~Abdelalim$^{\rm 49}$,
A.~Abdesselam$^{\rm 118}$,
O.~Abdinov$^{\rm 10}$,
B.~Abi$^{\rm 112}$,
M.~Abolins$^{\rm 88}$,
H.~Abramowicz$^{\rm 153}$,
H.~Abreu$^{\rm 115}$
% ...
\bigskip

% ...
$^{10}$ Institute of Physics, Azerbaijan Academy of Sciences, Baku, Azerbaijan\\
$^{11}$ Institut de F\'isica d'Altes Energies and Universitat Aut\`onoma  de Barcelona and ICREA, Barcelona, Spain\\
% ...
$^{48}$ Fakult\"{a}t f\"{u}r Mathematik und Physik, Albert-Ludwigs-Universit\"{a}t, Freiburg i.Br., Germany\\
$^{49}$ Section de Physique, Universit\'e de Gen\`eve, Geneva, Switzerland\\
% ...
$^{88}$ Department of Physics and Astronomy, Michigan State University, East Lansing MI, United States of America\\
% ...
$^{111}$ Homer L. Dodge Department of Physics and Astronomy, University of Oklahoma, Norman OK, United States of America\\
$^{112}$ Department of Physics, Oklahoma State University, Stillwater OK, United States of America\\
% ...
$^{115}$ LAL, Univ. Paris-Sud and CNRS/IN2P3, Orsay, France\\
% ...
$^{118}$ Department of Physics, Oxford University, Oxford, United Kingdom\\
% ...
$^{153}$ Raymond and Beverly Sackler School of Physics and Astronomy, Tel Aviv University, Tel Aviv, Israel\\
% ...
\end{flushleft}
